# Information paradox and black hole final state for fermions

Doyeol Ahn*

*Institute of Quantum Information Processing and Systems, University of Seoul, Seoul 130-743, Republic of Korea*

*-Abstract:* The black hole information paradox is the result of contradiction between Hawking's semi-classical argument, which dictates that the quantum coherence should be lost during the black hole evaporation and the fundamental principles of quantum mechanics – the evolution of pure states to pure states. For over three decades, this contradiction has been one of the major obstacles to the ultimate unification of quantum mechanics and general relativity. Recently, a final-state boundary condition inside the black hole was proposed to resolve this contradiction for bosons. However, no such a remedy exists for fermions yet even though Hawking effect for fermions has been studied for sometime. Here, I report that the black hole information paradox can be resolved for the fermions by imposing a final state boundary condition, which resembles local measurement with post selection. In this scenario, the evaporation can be seen as the post selection determined by random unitary transformation. It is also found that the evaporation processes strongly depends on the boundary condition at the event horizon. This approach may pave the way towards the unified theory for the resolution of information paradox and beyond.

*To whom correspondence should be addressed.
E-mail: dahn@uos.ac.kr ; davidahn@hitel.net

Hawking effect [1,2] on the information loss in black holes has been a serious challenge to modern physics because it contradicts the basic principles of quantum mechanics. Hawking's semi-classical argument predicts that a process of black hole formation and evaporation is not unitary [3]. On the other hand, there is an evidence in string theory that the formation and evaporation of black hole should be consistent with the basic principles of quantum mechanics [4]. Recently, a scheme based on the final-state boundary condition for the potential resolution of the black hole information paradox was proposed for bosons [4,5]. The essence of the proposal is to impose a unique final boundary condition at the black hole singularity such that no information is absorbed by the singularity. The final boundary state is maximally entangled states of the collapsing matter and infalling Hawking radiation. When a black hole evaporates, particles are created in entangled pairs with one falling into the black hole and the other radiated to infinity. The projection of final boundary state at the black hole singularity collapses the state into one associated with the collapsing matter and transfer the information to the outgoing Hawking radiation.

From a mathematical point of view, the Hawking effect arises from the logarithmic phase singularity of the wave functions [6] and it is known that Hawking radiation also exists for fermions as well [7-10]. However, a potential scheme to resolve the black hole information paradox for fermions does not exist to the best knowledge of the author, yet. In this Letter, I report that the black hole information paradox can be resolved for the fermions by imposing a final-state boundary condition. In this scenario, the evaporation can be seen as the measurement with post selection determined by random unitary transformation. It is also found that the evaporation processes strongly depends on the boundary condition at the event horizon.

We assume that the initial quantum state of the black hole belongs to a two dimensional Hilbert space $H_M$ and $|\psi\rangle_M$ be the initial quantum state of the collapsing matter. Let's assume that the orthonormal bases for $H_M$ are $\{|0\rangle_M, |1\rangle_M\}$ which correspond to different spins. The Hilbert space of fluctuations on the background spacetime for black hole formation and evaporation is separated into $H_{in}$ and $H_{out}$ which contain quantum states localized inside and outside the event horizon, respectively. In this proposal, the Unruh vacuum state $|\Phi_0\rangle_{in \otimes out}$ for fermions at the event horizon [11,12] belongs to $H_{in} \otimes H_{out}$:

$$|\Phi_0\rangle_{in\otimes out} = \sum_{i=0,1} A_n |i\rangle_{in} \otimes |i\rangle_{out}, \tag{1}$$

where $\{|i\rangle_{in}\}$ and $\{|i\rangle_{out}\}$ for $i = 0, 1$ are orthonormal bases for $H_{in}$ and $H_{out}$, respectively and $A_n = (1+\exp(-2\pi\Omega))^{-1/2}(-1)^n \exp(-n\pi\Omega)$. Here $\Omega$ is the energy spectrum of the particle. The final-state boundary condition (FBC) imposed at the singularity requires a maximally entangled quantum state in $H_M \otimes H_{in}$ which is called final boundary state and is postulated as

$$_{M\otimes in}\langle\Psi| = \sum_{n=0,1} A_n^* {}_M\langle n| \otimes {}_{in}\langle n|(S\otimes I), \tag{2}$$

where $S$ is a random unitary transformation. The initial matter state is given by $|\psi\rangle_M = c_o|0\rangle_M + c_1|1\rangle_M$. Then the initial matter state $|\psi\rangle_M$ evolves into a state in $H_M \otimes H_{in} \otimes H_{out}$, which is denoted by $|\Psi_0\rangle_{M\otimes in\otimes out} = |\psi\rangle_M \otimes |\Phi_0\rangle_{in\otimes out}$. Then the transformation from the quantum state of collapsing matter to the state of outgoing Hawking radiation is given by the following final state projection

$$\begin{aligned}|\phi_0\rangle_{out} &= {}_{M\otimes in}\langle\Psi\|\Psi_0\rangle_{M\otimes in\otimes out} \\ &= |A_0|^2 \left(c_{0M}\langle 0|S|0\rangle_M + c_{1M}\langle 0|S|1\rangle_M\right)|0\rangle_{out} + |A_1|^2\left(c_{0M}\langle 1|S|0\rangle_M + c_{1M}\langle 1|S|1\rangle_M\right)|1\rangle_{out} \\ &= \left(|A_0|^2|0\rangle_{out}\langle 0| + |A_1|^2|1\rangle_{out}\langle 1|\right)S\left(c_0|0\rangle_{out} + c_1|1\rangle_{out}\right) \\ &= PS|\psi\rangle_{out}\end{aligned} \tag{3}$$

where $P = |A_0|^2|0\rangle_{out}\langle 0| + |A_1|^2|1\rangle_{out}\langle 1|$ is a density operator ($|A_0|^2 + |A_1|^2 = 1$), which acts as a weighted measurement.

Previously, I showed that the change of boundary condition at the event horizon affect the evaporation process significantly for bosons [13]. It would be an interesting question to ask whether the black hole evaporation process for fermions will be affected by the boundary condition outside the event horizon. Now we consider the case of imposing Unruh excited state for fermions [11, 12] at the event horizon on the black hole evaporation problem. The Unruh excited state is obtained by applying the Bogoliubov transformation [7-12] on the Unruh vacuum state and is given by [11, 12]

$$|\Phi_1\rangle_{in\otimes out} = |0\rangle_{in} \otimes |1\rangle_{out}, \tag{4}$$

which is separable.

Then the initial matter state $|\psi\rangle_M$ evolves into a state $|\Psi_1\rangle_{M\otimes in\otimes out}$ in $H_M \otimes H_{in} \otimes H_{out}$, which is given by $|\Psi_1\rangle_{M\otimes in\otimes out} = |\psi\rangle_M \otimes |\Phi_1\rangle_{in\otimes out}$. The final state projection yields

$$\begin{aligned}|\phi_1\rangle_{out} &= {}_{M\otimes in}\langle\Psi\|\Psi_1\rangle_{M\otimes in\otimes out}\\ &= A_0^*{}_M\langle 0|S(c_0|0\rangle_M + c_1|1\rangle_M)|1\rangle_{out}\\ &= A_0^*|1\rangle_{out}\langle 0|S(c_0|0\rangle_{out} + c_1|1\rangle_{out})\\ &= TS|\psi\rangle_{out}\end{aligned} \quad (5)$$

where $T = A_0^*|1\rangle_{out}\langle 0|$.

Equations (3) and (5) show that the pure states always evolve to pure states under the black hole evaporation irrespective of the boundary condition at the event horizon, thus indicating that the black hole information paradox can be resolved for the case of fermions. The Unruh excited state corresponds to the state with a particle near the event horizon brought from the infinity. We note that $\|T\| < \|P\|$ where $\|\cdot\|$ is the operator norm. This result indicates that the presence of matter near the event horizon suppresses the transfer of information from the collapsing matter inside the black hole to the outgoing Hawking radiation and the evaporation of black hole.

It is interesting to note that the final state projections described by equations (3) and (5) resemble universal teleportation protocol [14], where entanglement plus local measurement and unitary transformation enables teleportation. In this protocol, the black hole evaporation is analogous to the measurement procedure done by Alice. The major difference exists, however, between the black hole evaporation and the quantum teleportation. In the latter, Bob needs a complete measurement results obtained by the classical channel to reconstruct the quantum state. On the other hand, the former doesn't need a classical channel because the evaporation is equivalent to sending the random unitary transformation as well as the state itself.

**Acknowledgements** This work was supported by the Korea Science and Engineering Foundation, the Korean Ministry of Science and Technology through the Creative Research Initiatives Program R16-1998-009001001-0 (2006). The author thanks M. S. Kim for drawing his attention to this issue.